\documentclass[english,american,superscriptaddress,prd,a4paper,nofootinbib,11pt]{revtex4-1}
\pdfoutput=1

\usepackage[T1]{fontenc}
\usepackage[latin9]{inputenc}
\usepackage{babel}
\usepackage{graphicx}
\usepackage[unicode=true,pdfusetitle,
 bookmarks=true,bookmarksnumbered=false,bookmarksopen=false,
 breaklinks=false,pdfborder={0 0 1},backref=false,colorlinks=false]
 {hyperref}
\usepackage[dvipsnames]{xcolor}

\usepackage{amsmath}

\usepackage{ulem}

\def\Mp{M_{\mathrm{pole}}}

\begin{document}

\title{Minimal Spin-one Isotriplet Dark Matter}

\author{Alexander Belyaev}
\email{a.belyaev@soton.ac.uk}
 \affiliation{School of Physics \& Astronomy, University of Southampton, Southampton SO17 1BJ, United Kingdom}
 
 \author{Giacomo Cacciapaglia}
 \email{cacciapa@ipnl.in2p3.fr}
 \affiliation{University of  Lyon, Université Lyon 1, CNRS/IN2P3, IPNL, F-69622, Villeurbanne, France.}

\author{James McKay}
 \email{jhmckay93@gmail.com}
 \affiliation{Department of Physics, Imperial College London, Blackett Laboratory, Prince Consort Road, London SW7 2AZ, UK}

\author{Dixon Marin }
\email{dixonjmp@gmail.com}
\affiliation{Departamento de Física and Centro Científico-Tecnológico de Valparaíso,
Universidad Técnica Federico Santa María, Casilla 110-V, Valparaíso,
Chile}

\author{Alfonso R. Zerwekh}
\email{alfonso.zerwekh@usm.cl}
\affiliation{Departamento de Física and Centro Científico-Tecnológico de Valparaíso,
Universidad Técnica Federico Santa María, Casilla 110-V, Valparaíso,
Chile}
\begin{abstract}
In this work we present a simple extension of the Standard Model
that contains, as the only new physics component, a massive spin-one matter field in the adjoint representation of $SU(2)_{L}$. In order
to be consistent with perturbative unitarity, the vector field
must be odd under a $Z_{2}$ symmetry. Radiative corrections make
the neutral component of the triplet ($V^{0}$) slightly lighter than
the charged ones. 
We show that $V^{0}$ can be the dark matter particle while satisfying
all current bounds if it has a mass between $2.8$ and $3.8$~TeV.
We  present the  current limit on the model parameter space from highly 
complementary  experimental constraints including
dark matter relic density measurement, 
dark matter direct and indirect detection searches, LHC data on
Higgs couplings to photons and LHC data on disappearing track searches. 
{We  show that the two-dimensional parameter space can be substantially
	covered by disappearing track searches at a future $100$~TeV hadron collider,
	which will probe DM mass upto about 1.2 TeV.}
\end{abstract}


\maketitle

\section{Introduction}
Our current microscopic understanding of Nature is based on the
Standard Model (SM) of particle physics. While the SM successfully describes the
properties of all known particles and their interactions, it is widely
believed to be incomplete.  The shortcomings of the SM are immediately evident when we consider the energy content of the Universe. 
Cosmological and astrophysical observations show that $26$\% of the mass-energy budget for the Universe is composed of an unknown dark matter (DM), while only $5$\% is accounted for by ordinary baryonic matter ($5$\%).  However, despite being a significant fraction of the the total mass-energy of the Universe and also having an important role in cosmological structure formation, the 
nature of DM remains unknown. One of the most popular expectations is that DM
could be made up of one or more types of weakly interacting massive particles
(WIMPs). Many WIMP candidates have been proposed,
including scalars~\cite{Deshpande:1977rw,LopezHonorez:2006gr,Hambye:2009pw,Belyaev:2016lok}, 
neutral fermions (the most notable example being neutralinos
appearing in supersymmetric models \cite{Goldberg:1983nd}), neutral gauge bosons (such as the Kaluza-Klein excitation of the photon \cite{Griest:1988ma,Cheng:2002ej,Servant:2002aq}
in models with universal extra dimensions) and light mesons in composite Higgs
models \cite{Frigerio:2012uc,Ma:2017vzm,Ballesteros:2017xeg}.  DM particles must be stable on at least cosmological time scales.
A popular means of ensuring stability is the introduction of discrete symmetries.  For example, the $Z_{2}$ symmetry appears in the form of R-parity in supersymmetry, T-parity in Little Higgs models and KK-parity in extra dimension (see for example Refs~\cite{Cacciapaglia:2009pa,Cacciapaglia:2016xty} for models with a geometrical origin on the KK parity and Refs~\cite{Cirelli:2005uq,Hambye:2009pw} for minimal cases with an accidental protection). 

In this work, we discuss the phenomenology of a novel WIMP candidate.  We extend the SM by
a massive spin-one \emph{matter} field in the adjoint representation
of $SU(2)_{L}$ (such that the new field is an isotriplet).  
{The main new feature of this model is that no additional states need to be included in the model to comply with 
gauge invariance and perturbative unitarity of the longitudinal vector polarisations. This allows us to define a truly
minimal model, defined in terms of 2 parameters: the mass of DM  and its coupling to the Higgs. Furthermore, radiative corrections
due to the electroweak gauge bosons ensure that the neutral component is the lightest one.}
While the case of spin-one DM has been studied in the literature before \cite{Gross:2015cwa,Chen:2014cbt,DiFranzo:2015nli,Kumar:2015wya,Duch:2015jta}), these models generally refer to gauge bosons related to new gauge symmetries under which only the dark sector is charged. In
our case, the new vector boson is not a gauge boson and is charged under the SM $SU(2)_L$ gauge group. 
In the absence of electroweak symmetry breaking (EWSB) the requirement of perturbative unitarity 
automatically imposes a $Z_{2}$ symmetry on this new vector field, 
thus preventing the lightest (neutral) component from decaying.  Therefore this $Z_{2}$ symmetry is
a requirement for the consistency of the model rather than simply introduced as a means to stabilise the DM candidate. The presence of the Higgs field and the associated EWSB reintroduces violation of perturbative unitarity.  However, this does not break the $Z_{2}$ parity at the level of renormalisable couplings. {Furthermore, the loss of perturbative unitarity can occur at a sufficiently high scale, orders of magnitde above the mass of the vector, so that no additional states and/or interactions relevant for the DM phenomenology need to be added to make the model self-consistent.}

The initial motivation for this work, originating from a previous
study \cite{Zerwekh:2012bf} performed by one of the authors, was the construction of a theory
containing a massive matter vector field coupled to a Yang-Mills field (with neither scalars nor symmetry breaking).
It has been shown in Ref.~\cite{Zerwekh:2012bf} that a theory containing a massive vector
boson in the adjoint representation, coupled to gauge bosons, is
 well behaved in the ultra-violet provided that: i) the self-interactions
of the new massive vector boson are governed by the same coupling as the gauge interactions, and ii) the new massive
vector boson is odd under a $Z_{2}$ symmetry. The latter is a consequence
of the absence of trilinear couplings among the new vectors.
Additionally, the resulting theory was shown to be BRST invariant.
In some sense, this construction defines a new kind of particle: a
vector boson which does not act as the carrier of an interaction but
plays the role of dark matter.

The model we will consider has two free parameters, the mass of the vector field and a coupling to the Higgs.  The viable parameter space is well constrained by experimental considerations, with an upper bound on the mass coming from a combination
of cosmology and DM direct detection experiments.  However, the remaining parameter space is difficult to test experimentally.
In this work, we evaluate the radiative corrections to the masses of 
the neutral ($V^{0}$) and charged ($V^\pm$)
components of the $SU(2)_L$ isotriplet.  These radiative corrections generate an essential mass splitting of $\approx 200$~MeV which renders the charged state to be short lived, thus making the isotriplet a viable DM candidate.  As a result of the small mass difference the dominant decay of the charged component is $V^{\pm}\to V^0 + \pi^\pm$, the width of which we have evaluated
using the pion effective theory. It turns out that the $V^{\pm}$ lifetime is on the order of 
$0.06$~ns, thus leading to disappearing charged track signatures from $V^{\pm}$ production,
with a characteristic length of $\approx 2$~cm only. This signature is potentially observable at the LHC 
and at future high-energy hadron colliders.
To find the available parameter space, we have exploited other theoretical and experimental constraints on the model parameter space,
such as:
i) constraints from perturbative unitarity loss that arise due to the $W^\pm$ mass and a non-vanishing coupling of the Higgs field; ii) constraints from LEP; iii) constraints from the Higgs LHC data; iv) constraints from the DM relic density and v) constraints from DM direct and indirect detection experiments.

The paper is organised as follows: in Section~\ref{sec:The-Model} we present the model and 
compute the radiative correction to the mass of the new particles.
In Section~\ref{sec:Results} we discuss the constraints on the two-dimensional parameter space of the
model, thus establishing the up-to-date bounds on the isotriplet mass and identifying the region
where the neutral component can account for the DM relic density. In Section~\ref{sec:pheno} we
estimate the reach at the LHC and future hadron collider by means of searches based on disappearing
tracks.  Finally we present our conclusions in Section~\ref{sec:Conclusions}.

\section{The Model} \label{sec:The-Model}

The introduction of massive vector fields into the SM without a companion scalar degree of freedom is generally not desirable.  Such theories are not renormalisable and violate perturbative unitarity.  However, there are some special cases where these problems can be avoided, such as in the Abelian field theories discussed in Ref.~\cite{Ruegg:2003ps}.  In Ref.~\cite{Zerwekh:2012bf} a non-Abelian gauge theory with a massive vector field ($V$) transforming homogeneously in the adjoint representation of the gauge group was studied. It was found that the theory preserves perturbative unitarity (at least at the tree-level) provided that the triple $V$ vertex is absent and that the quartic $V$ coupling is equal but opposite in sign to the quartic gauge boson coupling. The absence of the triple $V$ vertex makes the theory resemble the Abelian case and gives rise to an accidental $Z_2$ symmetry. In general, the quartic $V$ vertex would still contribute to unitarity violation for the scattering of the longitudinal component ($V_L V_L \rightarrow V_L V_L$). However, due to the fact that the quartic $V$ coupling has been linked to the gauge coupling, these dangerous terms are canceled out by a diagram that contains a gauge boson in the t-channel. In this sense, the gauge boson helps unitarising the $V_L$ scattering in the same way as the Higgs boson unitarises the $W_L$ scattering in the SM.   

In this work we apply the mechanisms described above to the SM supplemented by
a new massive vector boson in the adjoint representation of $SU(2)_L$. In other words, we construct an extension of the SM that includes
 a new massive isotriplet vector boson ($V_{\mu}$). By hypothesis, $V_{\mu}$ transforms homogeneously
 (\emph{i.e.} $V_{\mu}\rightarrow g_{L}^{\dagger}V_{\mu}g_{L}$ where
 $g_{L}\in SU(2)_{L}$). Additionally, we impose a $Z_2$ symmetry in order to avoid the triple $V$ vertex and we link the quartic $V$ coupling to the gauge coupling constant, as found in Ref.~\cite{Zerwekh:2012bf}. The resulting Lagrangian can be written as:
\selectlanguage{english}%
\begin{eqnarray}\label{eq:Lagrangian}
\mathcal{L} & = & \mathcal{L}_{SM}-Tr\left\{ D_{\mu}V_{\nu}D^{\mu}V^{\nu}\right\} +Tr\left\{ D_{\mu}V_{\nu}D^{\nu}V^{\mu}\right\} \nonumber \\
 &  & -\frac{g^{2}}{2}Tr\left\{  \left[V_{\mu},V_{\nu}\right]\left[V^{\mu},V^{\nu}\right]\right\} \label{eq:LagUnit}\\
 &  & -igTr\left\{ W_{\mu\nu}\left[V^{\mu},V^{\nu}\right]\right\} +\tilde{M}^{2}Tr\{V_{\nu}V^{\nu}\}\nonumber \\
 &  & +a\left(\Phi^{\dagger}\Phi\right)Tr\{V_{\nu}V^{\nu}\}\nonumber 
\end{eqnarray}
where $D_{\mu}=\partial_{\mu}-ig\left[W_{\mu},\;\right]$
is the usual $SU(2)_{L}$ covariant derivative in the adjoint representation 
and $\mathcal{L}_{SM}$ represents
the SM Lagrangian. The main difference with respect to the model in Ref.~\cite{Zerwekh:2012bf} is that the
$SU(2)_L$ symmetry is broken by the Higgs mechanism and the associated gauge bosons have mass. We thus allow for
a coupling of $V$ to the Higgs scalar field $\Phi$.
Due to the $Z_{2}$ symmetry the new vector boson
does not mix with the gauge bosons when the Higgs field acquires
a vacuum expectation value. In this sense, the EWSB is the same as in the SM. Consequently, the physical mass of the new vector bosons, $M_{V}$, is given by
\begin{equation}
M_{V}^{2}=\tilde{M}^{2}+\frac{1}{2}av^{2}\label{eq:MassV}
\end{equation}
where $v \sim 246$~GeV is the SM vacuum expectation value acquired by the Higgs field.

In Sections \ref{sec:Results} and \ref{sec:pheno} we will consider the phenomenological consequences of varying the two new parameters, $a$ and $M_V$.  We fix the parameters of the SM to current best-fit values, these are: $\alpha^{-1}_{\text{EM}}=129.950$ for the electromagnetic coupling, $s_W^2 = 0.23129$ for the Weinberg angle, $M_W = 80.385\,$GeV and $M_Z = 91.1876\,$GeV for the electroweak gauge boson masses and $125.5\,$GeV for the Higgs mass.  In our parameter space scans we consider values of $M_V$ from $100\,GeV$ up to $4\,$TeV and values of $a$ that are within the perturbative regime ($|a|\lesssim 4\pi$).  This mass range is sufficient to constrain the viable parameter space.  To begin with we will focus on the impact of the coupling to the Higgs, $a$, on the perturbative unitarity of the model in the following section.

\subsection{Higgs coupling and perturbative unitarity violation}

Although the model presented in Ref.~\cite{Zerwekh:2012bf} respects unitary at tree level, our extension of the SM has two sources of unitarity violation.  These result from the coupling of the heavy vector to the Higgs field and the fact that the SM $W$ and $Z$ bosons acquire mass via the Higgs itself.
 
The coupling to the Higgs scalar field, represented by the last term of the Lagrangian in Eq.~(\ref{eq:Lagrangian}), introduces a new Higgs $s$-channel contribution to $V_L V_L \rightarrow V_L V_L$ scattering. Together with the mass of the $W$ boson, which affects the $t$-channel, it reintroduces perturbative unitarity violation in the model.

In order to estimate the scale of unitarity violation, we study the  $V_L^+ V_L^- \rightarrow V_L^+ V_L^-$ process in the high energy limit. At tree level, seven diagrams contribute to this process in three topologies: the exchange of a Higgs boson, a $Z$-boson or a photon in the $s$ and the $t$ channels and a contact  $V_L^+ V_L^- V_L^+ V_L^-$ diagram. For large centre of mass energies the amplitude is
\begin{align}
-i{\cal M}=-\frac{\left(16 a^2 \sin^4(\theta_W) +3 e^4\right) \left(\cos(\theta)+1\right) M_W^2 s}{8 e^2\sin^2(\theta_W)M_V^4}.
\end{align}
We then expand the amplitude in partial waves. The most stringent unitarity constraint is obtained for $l=0$, for which the partial amplitude is
\begin{align}
a_0=\frac{1}{32\pi}\int_{-1}^1 {\cal M} \ d \!\cos(\theta)
\end{align}
Finally we apply the condition $|a_0|<1/2$ which gives the scale of unitarity violation to be
\begin{equation} \label{eq:lambda}
\Lambda \approx  \frac{8\sqrt{\pi} M_V^2}{\sqrt{4 a^2 v^2 + 3 g^2 M_W^2}}\,.
\end{equation}
As expected this scale goes to infinity when $v \to 0$ ($M_W \to 0$).

 \begin{figure}[htb]
\includegraphics[scale=0.53]{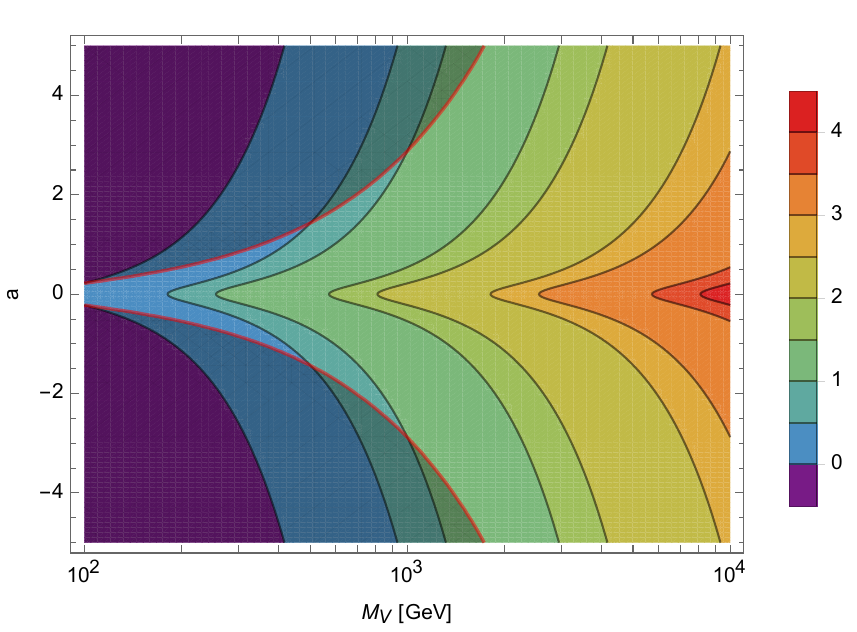}%
\includegraphics[scale=0.53]{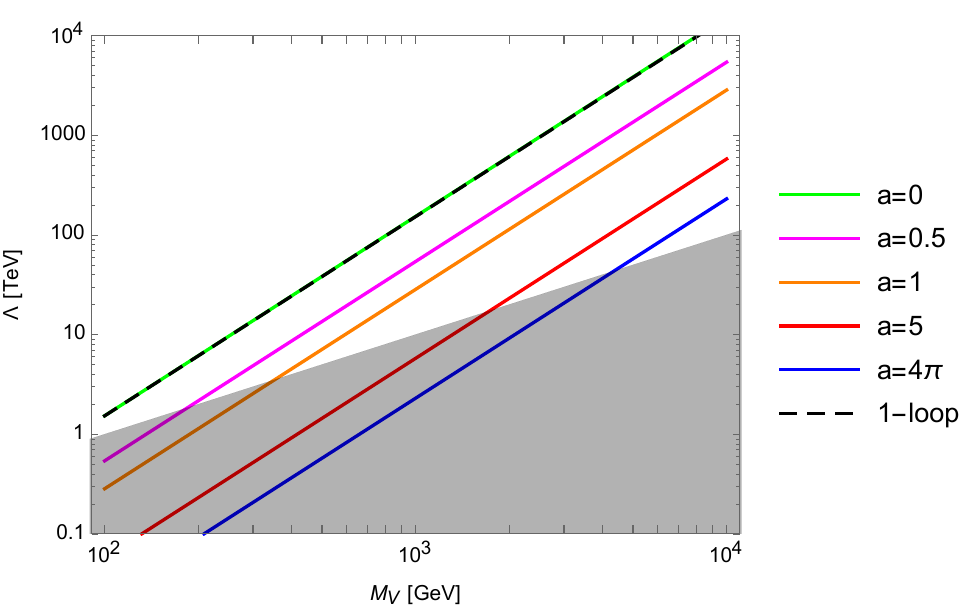}
  \caption{\label{fig:UnitarityScale}
  \textit{Left:} The expected scale of perturbative unitarity loss as a function of $M_V$ and $a$. 
Areas shaded by colors indicate the scale of unitarity violation $\Lambda$ according to Eq.\ref{eq:lambda}, while area limited by the red contour and shaded by grey color (for both left and right panels) corresponds to a loss of perturbative unitarity at a scale lower than $10\ M_V$.
 \textit{Right:} The expected scale of perturbative unitarity loss for the loop-induced $a$ (dashed), compared to the cases $a=0.$, $1.0$, $5.0$ and $4\pi$.}
 \end{figure}
 
In the left panel of Fig.~\ref{fig:UnitarityScale} we show the order of magnitude of $\Lambda$ as a function of $a$ and $M_V$. We see that for $M_V \gtrsim 1$~TeV (which is, as we will see below, the phenomenologically interesting region) and small $a$ we obtain $\Lambda \gtrsim 100$~TeV. This indicates that any new sector needed to complete the model and restore unitarity can exist at a high scale.  This makes the new sector irrelevant at the Fermi scale, thus rendering our model phenomenologically safe.
Small values for the coupling to the Higgs, $a$, imply higher values for the scale of unitarity violation.  For increasing $|a|$ the scale $\Lambda$ can become too close to the mass of the vector isotriplet, thus rendering the model inconsistent. The grey area in the plot indicates where $\Lambda < 10\ M_V$ and the theory is thus not reliable.

In principle it would seem convenient to simply fix $a=0$ in order to maximise the scale of unitarity violation, which would be given by
\begin{equation}
\Lambda(a=0)\approx 150~\mbox{TeV} \ \left( \frac{M_V}{1~\mbox{TeV}} \right)^2\,.
\end{equation}
However such a choice is not possible since if we eliminate the Higgs-$V$ coupling term at tree level it will be reintroduced by quantum effects at higher loop order.
From the Lagrangian in Eq.~(\ref{eq:Lagrangian}), we can see that loops of the vector isotriplet and the SM $SU(2)_L$ gauge bosons contribute to generating an effective coupling between the Higgs field and the triplet. By explicit computation we find that logarithmic divergences arise at one-loop level which require a non-zero counter-term from the $a$ coupling.  We can thus estimate the size of $a$ to be:
\begin{equation}
a^{\rm 1-loop} = - 3 \frac{\alpha^2}{\sin \theta_W^4} \ln \frac{\Lambda}{M_V} \approx - 0.0037\; \ln \frac{\Lambda}{M_V}\,,
\end{equation}
where $\Lambda$ is the cut-off of the theory (where we assume $a$ to vanish), and we fix the renormalisation scale to $M_V$. 
We identify $\Lambda$ with the scale of perturbative unitarity loss, and plot this using Eq.~(\ref{eq:lambda}) in the right panel of Fig.~\ref{fig:UnitarityScale}.  This leads to a value of the loop-induced couplings that is mildly dependent on $M_V$, with $a^{\rm 1-loop} (M_V = 1~\mbox{TeV}) \approx - 0.02$.
Fig.~\ref{fig:UnitarityScale} shows that the impact of the loop induced coupling is always negligible compared to the effect of the $W$ mass, as the dashed line overlaps with the $a=0$ curve.  However, as shown in the plot large values of the coupling will significantly lower the expected scale of perturbative unitarity loss.
In addition, for fixed $a$ the consistency of the theory requires a lower bound on the mass of the vector field, which ranges between $100$~GeV and $1$~TeV.
For masses above $\sim 2\,$TeV, perturbative unitarity violation occurs at sufficiently high scales as long as the coupling $a$ remains perturbative.

{The simplest UV completion of the theory would be 
the introduction of the additional scalar doublet which would generate the mass for the new vector doublet  via spontaneous symmetry breaking mechanism, while vector doublet would take a role of  a gauge field of the symmetry. Eventually this  completion predicts the new heavy scalar state with mass between $M_V$ and  $\Lambda$, the perturbative unitarity violation scale which we have estimated  for our minimal model.}


\def\Mv{M_V}
\newcommand{\MSBar}{\overline{MS}}
\newcommand{\MSbar}{$\MSBar$ }
\subsection{The radiatively induced mass splitting}\label{sec:The-mass-splitting}

Another important feature of the vector model is the radiatively induced mass splitting between the components.  At the tree-level, the neutral and charged isotriplet components are degenerate, having the same mass $M_V$.  Fortunately, radiative corrections due to electroweak interactions at the one-loop order and above induce a mass splitting, making the neutral boson lighter than the charged ones.
This mechanism is an essential feature of a number of DM theories, including triplet scalar DM \cite{FileviezPerez:2008bj,Fischer:2011zz,Araki:2011hm,Araki:2010zz}, Minimal DM \cite{Cirelli:2005uq,Yamada:2009ve} and the wino-limit of $R$-parity conserving supersymmetry (where the rest of the supersymmetric spectrum is decoupled and a pure wino-like neutralino is a potential DM candidate) \cite{Cheng:1998hc,Feng:1999fu,Ibe:2012sx,McKay:2017xlc}.  This mass splitting is typically found to be on the $\sim 100$\,MeV scale for DM masses greater than $\sim100$\,GeV.  

We define the physical masses for the charged and neutral components of the vector multiplet as $\Mp^+$ and $\Mp^0$ respectively. These pole masses are given by $p^2$ satisfying
\begin{equation}
p^2  = M^2_V  - \Sigma^i_{V}(p^2)\,,\label{eqn:pole}
\end{equation}
where $\Sigma^i_V$ is the real and transverse part of the self-energy for the charged ($i=+$) or neutral ($i=0$) component of the multiplet.  Equivalently, up to one-loop order, the pole masses are given by
\begin{equation}
\Mp^i = \sqrt{M_V^2-\Sigma^i_V(M_V^2)}\,.\label{eqn:pole_mass}
\end{equation}
The mass splitting between the physical masses of the charged and neutral components can be written by expanding Eq.~(\ref{eqn:pole_mass})
\begin{equation}\
\Mp^i
=M_V \sqrt{1-\frac{\Sigma^i(M_V^2)}{M_V^2}}
=M_V \sum_{n=0}^\infty  (-1)^{n} {\frac{1}{2} \choose n}   \left(\frac{\Sigma^i(M_V^2)}{M_V^2}\right)^n\,, \label{eqn:Mpole_approximation}
\end{equation}
and taking the difference
\begin{eqnarray}
\Delta M = \Mp^+-\Mp^0 = \Mv\sum_{n=1}^\infty  (-1)^{n} {\frac{1}{2} \choose n}  \left[ \left(\frac{\Sigma^+(\Mv^2)}{\Mv^2}\right)^n-  \left(\frac{\Sigma^0(\Mv^2)}{\Mv^2}\right)^n\right]\,.
\label{eqn:deltam}
\end{eqnarray}
For a consistent one-loop result we truncate the expansion to the first term, which appears at order $g^2$ in the gauge couplings.

We compute the self-energies in the Feynman-'t Hooft gauge at one-loop order.  The one-loop, $\mathcal{O}(g^2)$, mass splitting is obtained from the first term in the expansion in Eq.~(\ref{eqn:deltam}),
\begin{align}
\Delta M = \frac{g^2 }{ 12\, (16\pi)^2 \Mv^3}
&\left[ f(M_W) + g(M_W) - c_W^2  \left(f(M_Z)-   g(M_Z) \right)\right.\nonumber\\
 &\left.+ 5(M_W^2-c_W^2M_Z^2) (A(\Mv)-2\Mv^2) + 30 s_W^2 \Mv^4 B(\Mv,0) \right]\label{eqn:one_loop_deltam}
\end{align}
where
\begin{eqnarray}
f(x) &=& -(30\Mv^4 + 26 \Mv^2 x^2-5 x^4)B(\Mv , x)\\
g(x) & =& (12\Mv^2-5x^2) A(x)
\end{eqnarray}
and $A$ and $B$ are defined in Eqs.~(\ref{A0 def}) and (\ref{B0 def}) respectively.  

\begin{figure}
\centering
\includegraphics[width=0.5\textwidth]{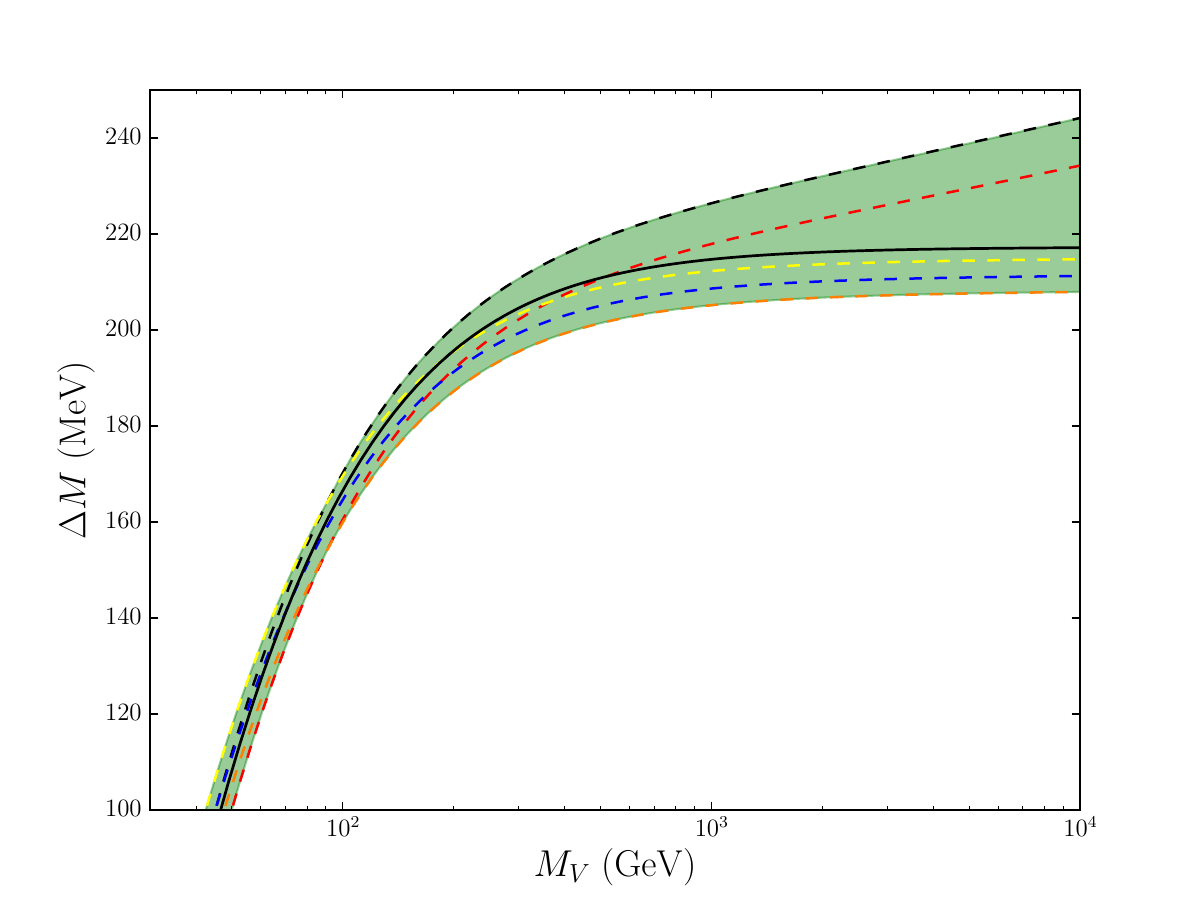}
\includegraphics[width=0.9\textwidth]{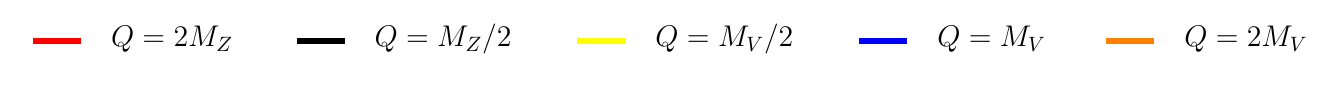}
\caption{The one-loop radiatively induced mass splitting between the charged and neutral components of the vector DM.  The solid lines represent $\Delta M$ computed at fixed values of the renormalisation scale $Q$.  The shaded green band indicates the range of values obtained by varying $Q$ continuously between $\min\{M_V/2,M_Z/2\}$ and $\max\{2M_V,2M_Z\}$ and thus constitutes an estimate of the uncertainty on $\Delta M$. The solid black line is the one-loop mass splitting, Eq.~(\ref{eqn:one_loop_deltam}), with all higher order terms truncated.\label{fig:mass_splitting}}
\end{figure}

To evaluate the mass splitting for $\Mv \gg M_W,M_Z$, we use the limits from Eqs\ (\ref{eqn:B0_limit}) and (\ref{eqn:A0_def}), which gives
\begin{equation}
\Delta M = \frac{5 g_W^2 (M_W-c_W^2 M_Z)  }{32 \pi } \approx 217.3 \,\rm{MeV}.
\label{eqn:mass_splitting}
\end{equation}
Like in the fermionic case, this result is independent of $\Mv$ in the large $\Mv$ limit.  A plot of the full expression in Eq.~(\ref{eqn:one_loop_deltam}) as a function of $M_V$ is presented in Fig.~\ref{fig:mass_splitting} (black solid curve), where we see that the asymptotic constant value is reached for masses above $\sim$$\,500$~GeV.

The non-truncated mass splitting, Eq.~(\ref{eqn:deltam}), contains higher order terms.  The sensitivity of these higher order terms to the renormalisation scale can be used to give a naive estimate of the theoretical uncertainty in the one-loop result. The dependence on $Q$ enters both via the input parameters\footnote{For the one-loop mass splitting we need only compute the running \MSbar coupling, which we take as $e_{\rm{SM}}(m_Z) = 0.3134$ and renormalise using the one-loop SM renormalisation group equation.  Any matching to the vector mass is of higher order and thus can be neglected.} and via explicit logarithms. In the case of a fermionic multiplet in the large mass limit, only the former $Q$ dependence appears \cite{Ibe:2012sx,McKay:2017rjs} thanks to a cancellation of all scale-dependent logarithms between the neutral and charged self-energies: this is due to the fact that the fermionic pole mass is linear in the self-energies, which share the same dependence on the logarithms. The vector case is different because it is the mass squared that depends linearly on the self-energies, thus a cancellation of the explicit logarithms occurs in $(M^+_{\rm{pole}})^2-(M^0_{\rm{pole}})^2$, but not in $\Delta M = M^+_{\rm{pole}}-M^0_{\rm{pole}}$.

In fact, as seen in Eq.~(\ref{eqn:deltam_series}), the next-to-leading terms in the series ($\mathcal{O}(g^n)$ for $n>2$) explicitly contain the term $\log(M_V/Q)$.
The dotted lines in Fig.~\ref{fig:mass_splitting} show the results for $Q = M_Z/2, \, 2M_Z,\, M_V/2,\, \Mv$ and $2\Mv$. This allows us to identify the theoretical uncertainty with the green region, and estimate the error in the range $5 - 10\,$\%.  However, this is only a naive estimate for the uncertainty in the one-loop result, and should be considered along with an estimate of the magnitude of missing two-loop corrections, such as that performed for a fermionic multiplet in \cite{Ibe:2012sx,McKay:2017xlc}.

\subsection{Electroweak precision tests}

The oblique formalism of Peskin and Takeuchi \cite{Peskin:1991sw} is a convenient way to parameterise the effect of new physics on electroweak radiative corrections.  
In an iso-spin conserving theory, $T$ and $U$ are identically zero, thus we only need compute $S$ which is given by
\begin{align}
S = \frac{4s_W^2c_W^2}{\alpha} \left[\Pi'_{ZZ}(0)-\frac{c_W^2-s_W^2}{s_Wc_W}\Pi'_{Z\gamma}(0)-\Pi'_{\gamma\gamma}(0)\right]\,, \label{eqn:S}
\end{align}
where the $\Pi_{ij}(p^2)$ are the one-loop contributions to the self energies from the new vector field, for the processes $Z\rightarrow Z$, $\gamma\rightarrow \gamma$ and $\gamma\rightarrow Z$ given by $ij = ZZ$, $\gamma\gamma$ and $\gamma Z$ respectively.  The prime denotes a derivative with respect to the external momentum squared ($p^2$).  With divergences appropriately subtracted, these contributions are found to be
\begin{align}
\Pi_{\gamma\gamma} = \frac{e^2}{18 (16\pi^2)^2} \left( -2 (-21 A(\Mv)+2\Mv^2+p^2)-3B(\Mv,\Mv)(32\Mv^2+19p^2)\right)
\end{align}
and $\tan^2\theta_W\Pi_{Z\gamma} = \tan\theta_W\Pi_{ZZ} = \Pi_{\gamma\gamma}$.  Taking the derivative of each of these and using Eq.~(\ref{eqn:S}) we find that $S=0$.
Non-vanishing corrections to the oblique parameters will, therefore, only arise at two-loop level, thus leading to very mild bounds on the mass of the new vector isotriplet.

\section{Numerical results\label{sec:Results}}

We now turn to analysing the phenomenology of this vector DM model, with particular attention to the role of the neutral component as a DM candidate.

\subsection{Relic Density, Direct and Indirect Searches}

\begin{figure}
\includegraphics[width=0.7\textwidth]{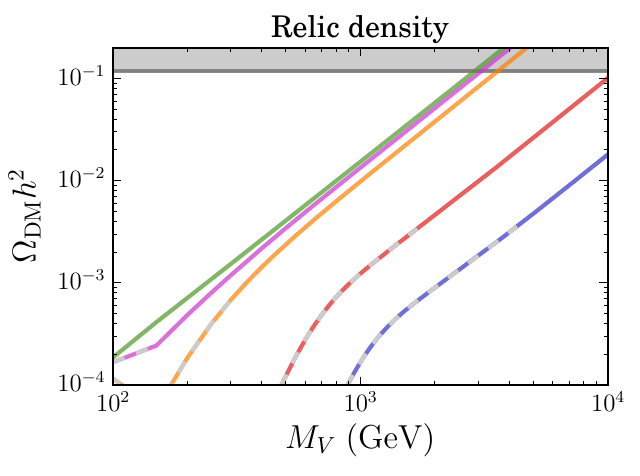}\vspace{-0.3cm}\raisebox{2.4cm}{\includegraphics[width=0.3\textwidth]{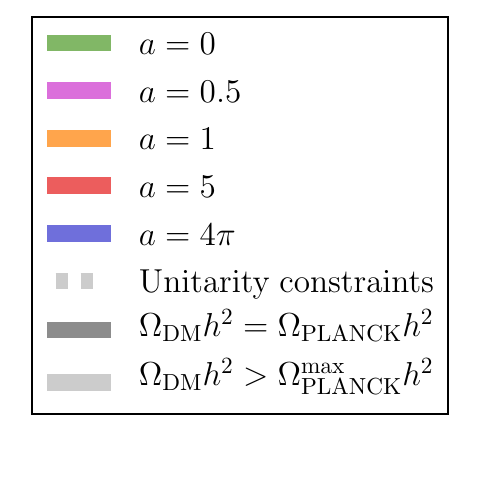}}
\caption{\label{fig:Relic-density}Thermal relic density for  $V^{0}$ vs $M_{V}$ for
various values of $a$. Perturbative unitarity loss at $\Lambda < 10 M_V$ occurs where grey dashing overlays the relic density
lines. The grey horizontal band corresponds to the range measured
 by Planck, $\Omega_{DM}h^{2}=0.1186\pm0.0020$, and the light grey region indicates $\Omega_{DM}h^{2}>0.1206$.}
\end{figure}

One of the most interesting aspects of this model is the cosmological consequences.  For the analysis of these consequences we use the micrOMEGAs package \cite{Belanger:2013oya}. 
We start by simply computing the relic density, $\Omega_{DM}$, of $V^{0}$
for representative values of $M_{V}$ and $a$. Our results are shown in Fig.~\ref{fig:Relic-density}.  The dark coloured grey horizontal band indicates the region where DM relic density is within one sigma of the value measured by Planck \cite{Adam:2015rua}, $\Omega_{\text{PLANCK}}h^{2}=0.1186\pm0.0020$.  The region above this, shaded light grey, indicates where the relic density exceeds the Planck measured value by more than one sigma (and is thus ruled out).  The various curves are computed for fixed values of $a$. We see
that the DM thermal relic abundance can always match the Planck result
for sufficiently large values of the mass $M_V$, with the lowest value
$M_V \approx 2.85\,$TeV
attained for $a=0$.  For increasing $a$ the mass of the DM increases into the TeV range, with a maximal value of several tens of TeV
attained at the value of $a=4\pi$.
We also remark that lower values of the mass are also allowed with $V^0$
constituting only a fraction of the DM relic density, down to the mass value
where perturbative unitarity loss occurs at a too low scale, as indicated
by the dashed portion of the curves,
where $M_V>\Lambda/10$, with
$\Lambda$ given in Eq.~\eqref{eq:lambda}.

\begin{figure}
\includegraphics[width=0.7\textwidth]{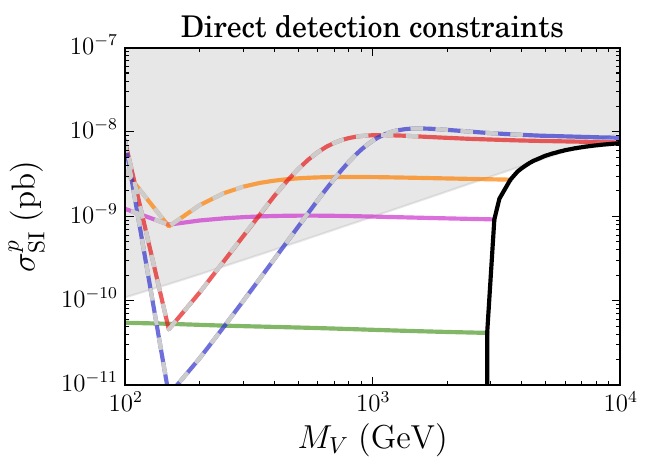}\vspace{-0.3cm}\raisebox{2.4cm}{\includegraphics[width=0.3\textwidth]{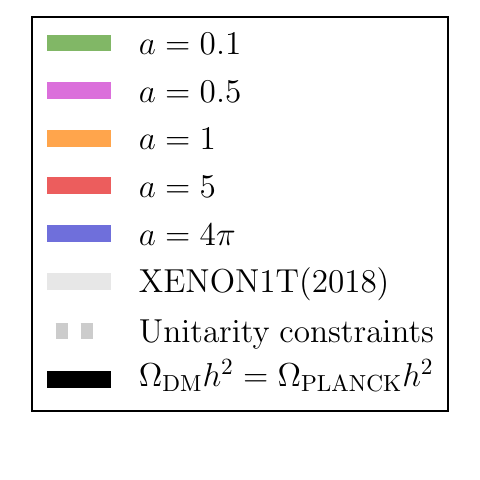}}
\caption{\label{fig:DirectSearches} Spin-independent cross-section for $V^{0}$-nucleon
elastic scattering as a function of $M_{V}$ and for representative
values of $a$. The cross-section has been rescaled to take into account the actual
thermal relic abundance. The continuous black curve represents
the elastic cross-section computed with the values of $M_{V}$ and
$a$ that saturate the measured DM relic density. The grey dashing
highlights the parameter space where perturbative unitarity loss occurs
at too low scale. }
\end{figure}

The vector DM can interact with SM nucleons via the Higgs coupling $a$, this allows us to place bounds on the values this coupling may take using limits from direct experiments.  There are also loop induced couplings to quarks and leptons, via the electroweak sector, which are sufficiently small to not be of interest in this study.\footnote{The effective coupling $V^0 V^0 \bar{q} q$ is also suppressed by the mass of the quark due to the chirality of the coupling.} Finally, a trilinear coupling with the Z-boson ($V^0V^0Z$) is absent in our model.  We thus compute the spin-independent scattering cross-section on protons, $\sigma_{\rm SI}$, for
various values of $a$, as shown in Fig.~\ref{fig:DirectSearches}, and compare them
with the 2018 results from the XENON1T experiment~\cite{Aprile:2017iyp,Aprile:2018dbl}.  The region of the parameter space excluded by the perturbative unitarity constraint is indicated by the overlaid grey dashed lines on these lines of fixed $a$.  The XENON1T results give significant improvements with respect to LUX~\cite{Akerib:2016vxi} and PandaX II~\cite{Cui:2017nnn}, so we do not consider the latter in this study.  

When the thermal relic abundance of vector DM is less than that measured by the Planck experiment, we rescale the cross-section appropriately to account for this.  The grey line indicates points for which the vector constitutes one hundred percent of the relic abundance of DM.  We see that the latest XENON1T results already exclude values of $|a| > 1$, even for masses for which the relic abundance is small.

Therefore the DM mass in the surviving parameter space is limited values less than $\sim4\,$TeV. There is also a surviving
region for $|a| \gtrsim 5$ and $M_V \gtrsim 8.5\,$TeV.  However, due to the large couplings involved
in this region our leading order calculations will be unreliable due to large missing higher order corrections, so we currently do not consider this region further.
The dip in $\sigma_{\rm SI}$ at low DM mass ($M_V \approx M_h$) and large $|a|$ is due to resonant Higgs annihilation.  However, in this region we find that the scale of perturbative unitarity loss is lower than lower than our constraint ($\Lambda < 10M_V$).
The next update of the XENON detector~\cite{Aprile:2018dbl} will improve the constraining power by at least one order of magnitude in the scattering cross-section, thus bringing the bound to $|a| \approx 0.3$.  Future experiments like LUX-ZEPLIN~\cite{Akerib:2018lyp} and PandaX 4T~\cite{Zhang:2018xdp} will improve the bound by a further order of magnitude, thus probing down to $|a| \approx 0.1$.

\begin{figure}
\includegraphics[width=0.7\textwidth]{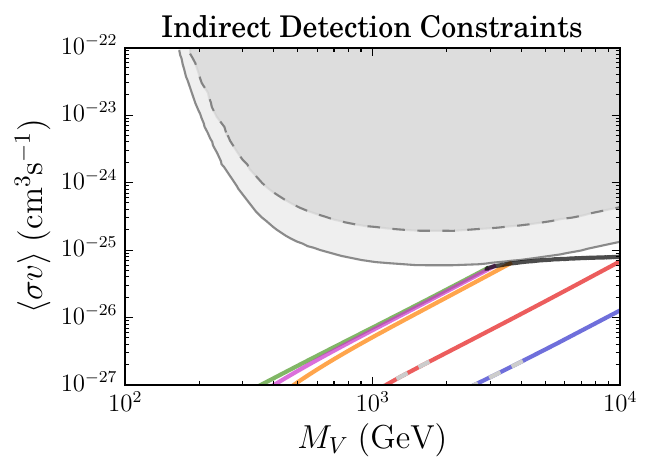}\vspace{-0.3cm}\raisebox{2.4cm}{\includegraphics[width=0.3\textwidth]{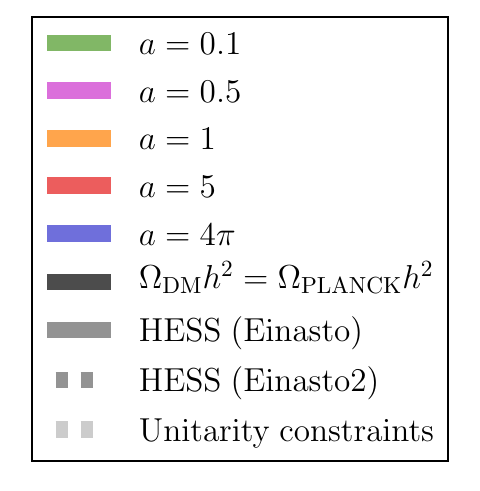}}
\caption{\label{fig:IndirectSearches} Annihilation cross-section of $V^0$ compared to the most recent bounds
from HESS in the $W^+ W^-$ channel. The two bands consist of two different choices of the DM profile
model (Einasto), showing the sensitivity of the bound to the DM distribution at the centre of the Galaxy. }
\end{figure}

The main annihilation channel for DM in this model is into a W boson pair, with the final states $ZZ$ and $hh$ also produced via the coupling $a$. Thus the model can be probed by indirect detection experiments looking for cosmic rays generated by these decay products. The most sensitive
channel will be into photons emitted by the decay products of the $W^\pm$, $Z$ and Higgs. As the
photon fluxes are very similar for the three final states in our model~\cite{Cirelli:2010xx}, we compare the velocity averaged
cross-section of DM with the HESS results for the $W^+ W^-$ channel~\cite{Abdallah:2016ygi}. The result is shown 
in Fig.~\ref{fig:IndirectSearches}, with the cross-section rescaled appropriately when the relic abundance is less than that measured by Planck. 
Note that our annihilation cross-section is higher than the usual one considered
to saturate the thermal relic abundance because we have co-annihilation with the charged
vector $V^\pm$. As a result, when saturating the Planck relic density (black line), the lower mass values for $a\simeq 0$, which are not detectable in direct detection experiments, 
are close to the exclusion deriving from the most optimistic
DM profile. Bounds that are less sensitive to the details of the DM profile in the centre of the Galaxy can be obtained
by observing light emitted by dwarf spheroidal galaxies in experiments like VERITAS~\cite{Zitzer:2015eqa} and
MAGIC~\cite{Ahnen:2016qkx}. However, these bounds are weaker by at least one order of magnitude.  Profile-independent (and thus more robust) bounds can be  obtained by detecting distortions of the cosmic microwave background due to photons
injected by DM annihilation in the early Universe~\cite{Slatyer:2015jla}. However, for heavy DM masses above $1\,$TeV
such bounds are greater than $10^{-24}$~cm$^3$/s and are thus too weak to constrain this model.

\subsection{Contribution to $h\rightarrow\gamma\gamma$}

The charged components of the vector isotriplet contribute to the Higgs boson decay into two photons via loop effects.  The Lorentz structure
of this contribution is exactly the same as that of the $W^{\pm}$ bosons.  This fact simplifies the computation of the partial
decay width for this channel. The result, expressed in terms of the partial decay width, is
\begin{eqnarray} \label{eq:HtoAA}
\Gamma\left(h\rightarrow\gamma\gamma\right)  =  \frac{\alpha^{2}\,M_{h}^{3}}{256\pi^{3}v^{2}} \left[ N_{c}Q_{t}^{2}F_{1/2}\left(x_{t}\right) + F_{1}\left(x_{W}\right)
 +\frac{a}{2}\ \left(\frac{v}{M_{V}}\right)^{2}F_{1}\left(x_{V}\right) \right]^{2}\,,
\end{eqnarray}
where $\alpha$ is the electromagnetic fine-structure constant,
 $M_{h}$ is the mass of the Higgs boson and $x_{i}=m_{h}^{2}/4M_{i}^{2}$. 
 The first term in Eq.
(\ref{eq:HtoAA}) represents the contribution of the top quark ($N_{c}$
and $Q_{t}$ are the number of colours and the electric charge of the
top quark, respectively) while the second term originates from the $W^{\pm}$
boson loops. The third term is due to the presence of the $V^{\pm}$
bosons. The functions $F_{1/2}$ and $F_{1}$ are the one-loop factors for particles
of spin-half and spin-one respectively, and are defined in Ref.~\cite{Spira:1995rr}.
In this model, only $\Gamma\left(h\rightarrow\gamma\gamma\right)$ is sizeably affected by the presence of the vector triplet
as other Higgs couplings will only receive next-to-leading order corrections.
Therefore we can compare the modification of the partial width,
which we embed in the ratio
\begin{equation}
R_{\gamma\gamma}=\frac{\Gamma\left(h\rightarrow\gamma\gamma\right)}{\Gamma\left(h\rightarrow\gamma\gamma\right)_{\text{SM}}}
\end{equation}
where $\Gamma\left(h\rightarrow\gamma\gamma\right)_{\text{SM}}$ is the partial
width predicted in the SM, to the signal strength of the diphoton channel measured at the LHC. 
The most recent results from ATLAS~\cite{Aaboud:2018xdt} and CMS~\cite{Sirunyan:2018ouh}, obtained from an integrated luminosity of approximately
$36\ \mbox{fb}^{-1}$ respectively at $13\,$TeV, give
\begin{equation}
R_{\gamma\gamma}^{\rm ATLAS} = 0.99 \pm 0.14\,, \qquad R_{\gamma\gamma}^{\rm CMS} = 1.18^{+0.17}_{-0.12}\,.
\end{equation} 
The Run-II results significantly improve over Run-I~\cite{Khachatryan:2016vau} by nearly a factor of two reduction in the error.
\begin{figure}
\includegraphics[scale=0.7]{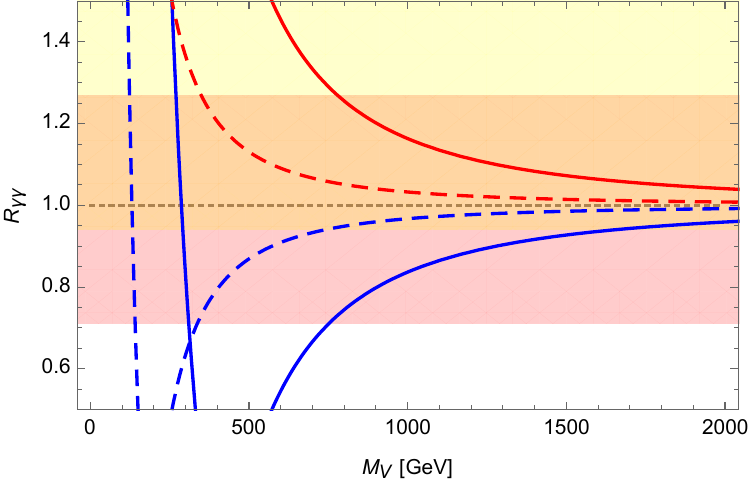}
\caption{\label{fig:HtoAA} Contribution of the new vector isotriplet to the $h\rightarrow\gamma\gamma$
decay channel for $a = \pm 1$ (dashed) and $a = \pm 5$ (solid). The colour code uses red for positive values and blue for negative. 
The coloured bands are the experimentally allowed regions at $95\%$ CL from ATLAS (pink) and CMS (yellow), while the orange band shows the overlap.}
\end{figure}

To illustrate the numerical impact of the
vector isotriplet on the Higgs to diphoton production rates at the LHC, in Fig.~\ref{fig:HtoAA}
we plot $R_{\gamma\gamma}$ as a function of $M_{V}$
for values of $a = \pm 1$ and $\pm 5$. The coloured bands are the experimentally
allowed regions at $95\%$ CL level ($2\sigma$) from ATLAS (pink) and CMS (yellow). 
The orange overlapping region is a very conservative estimate of a combination of the
two experiments, while a true combination would lead to a significantly larger band~\cite{Khachatryan:2016vau}.
For masses larger than about $1.5\,$TeV, the computed values of $R_{\gamma\gamma}$ are 
consistent with reported measurements.
We also remark that a cancellation for negative values of $a$ occurs, as seen in the blue lines of Fig.~\ref{fig:HtoAA}. 
This is due to the cancellation between the top and $W$ loops in the SM, which have opposite signs.  A negative 
$a$ would add to the top contribution, thus eventually overshooting the $W$ loop contribution at large $|a|$
or small masses.
In the region relevant for thermal DM, with masses above $2.5\,$TeV, the contribution is smaller than $5\%$ for $|a|<5$.  Therefore it would require a percent level measurement of the diphoton signal strength to be able to probe this model.

\subsection{Results from the two-dimensional scan}

\begin{figure}[htb]
\centering
\includegraphics[width=0.5\textwidth]{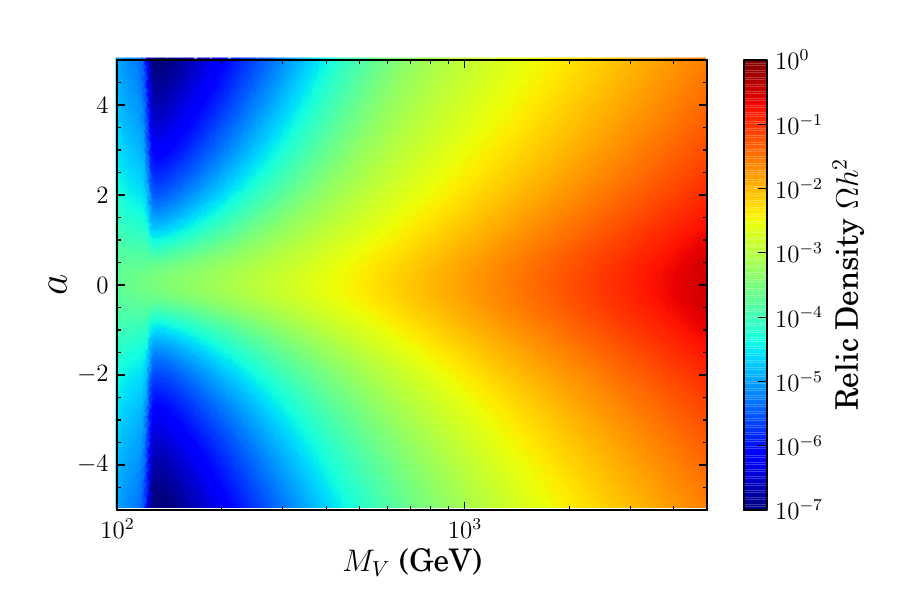}%
\includegraphics[width=0.5\textwidth]{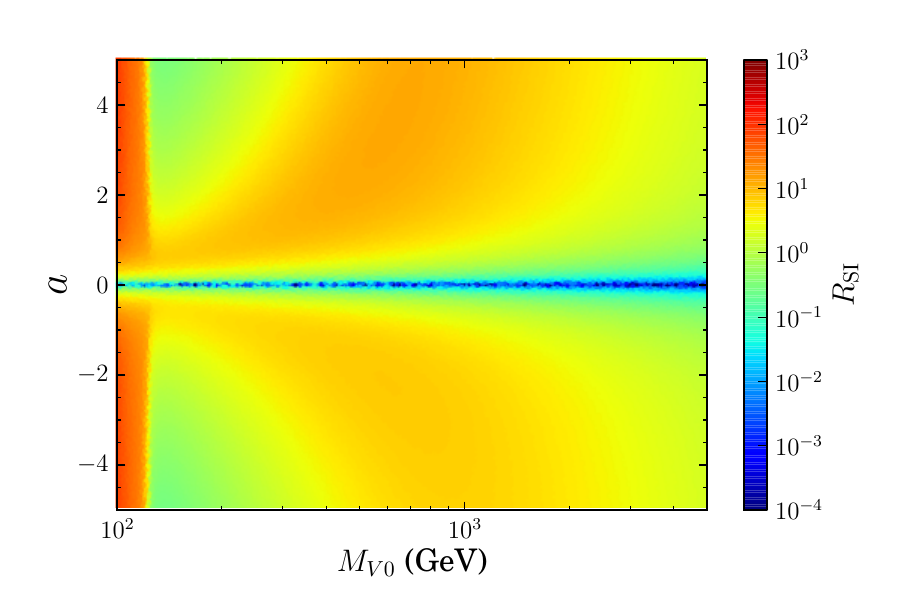}\\
\includegraphics[width=0.5\textwidth]{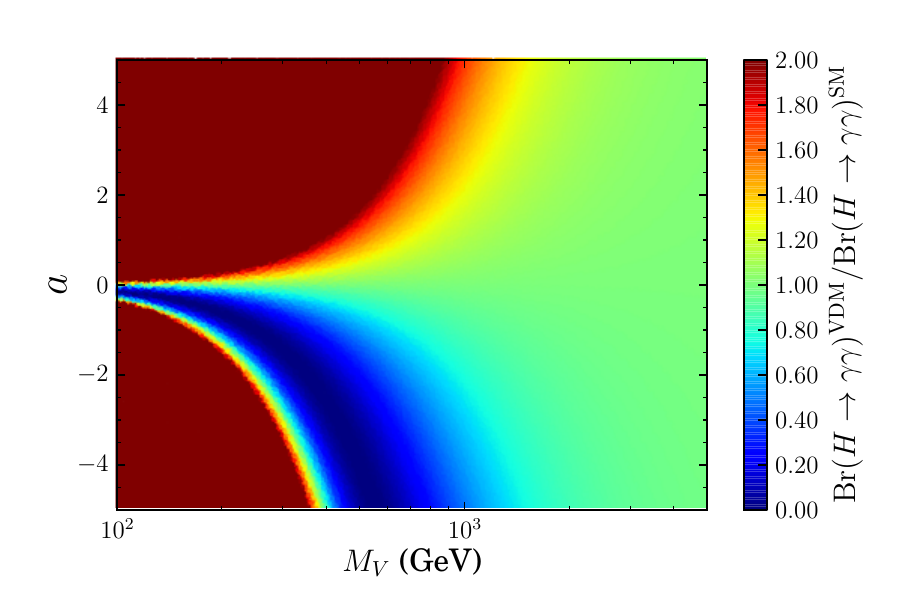}
\caption{\label{fig:scan1} Contribution of the vector isotriplet to the DM relic abundance (\textit{top-left}), spin-independent direct detection cross-section in terms of the ratio in Eq.~(\ref{eqn:RSI}) (\textit{top-right}) and $R_{\gamma \gamma}$ (\textit{bottom}), as a function of $M_V$ and $a$.}
\end{figure}

In Fig.~\ref{fig:scan1} we show the effect of the presence of the vector isotriplet for three observables.  In the top left panel we show the DM relic density.
In the top right panel we plot the ratio
\begin{align}
R_{\text{SI}} = \frac{\sigma^p_{\text{SI}}}{\sigma^p_{\text{SI, XENON1T}}} \frac{\Omega_{\text{DM}}}{\Omega_{\text{PLANCK}}} \label{eqn:RSI}
\end{align}
where $\sigma^p_{\text{SI}}$ is the spin-independent scattering cross-section of the vector dark matter and $\sigma^p_{\text{SI, XENON1T}}$ is a parameterisation of the central value of the XENON1T limit (given by lower boundary of the grey region in Figure \ref{fig:DirectSearches}).  In the lower panel we present the Higgs partial width into two photons.  Note that only the Higgs partial width strongly depends on the sign of $a$. For negative $a$, the isotriplet loop interferes 
destructively with the $W$ loops, thus reducing the partial decay width of the Higgs in two photons. When the coupling $a$
becomes large, the $V^+$ loop overwhelms the SM contributions. Thus, there is a thin strip for negative $a$ for which 
the result accidentally matches the SM result.

\begin{figure}[htb]
\centering
\scalebox{1.1}{\hspace{-0.5cm}\includegraphics[height=0.25\textwidth]{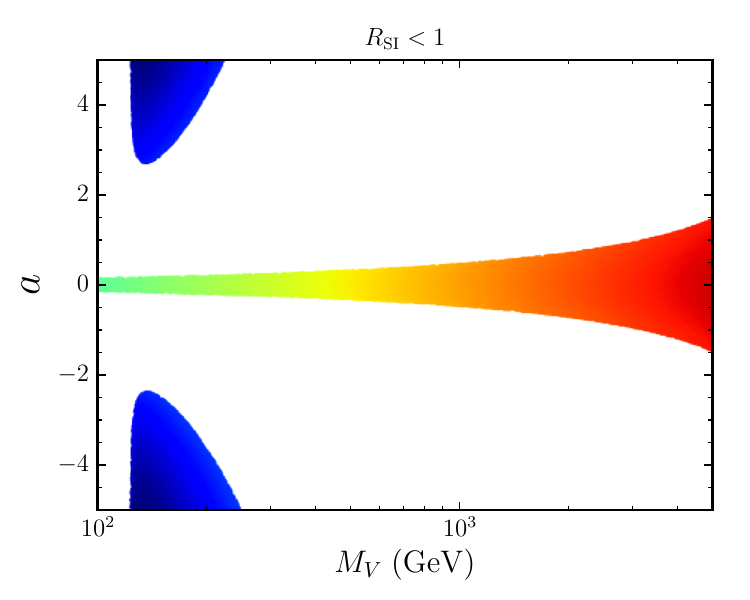}\hspace{-0.255cm}\includegraphics[height=0.25\textwidth]{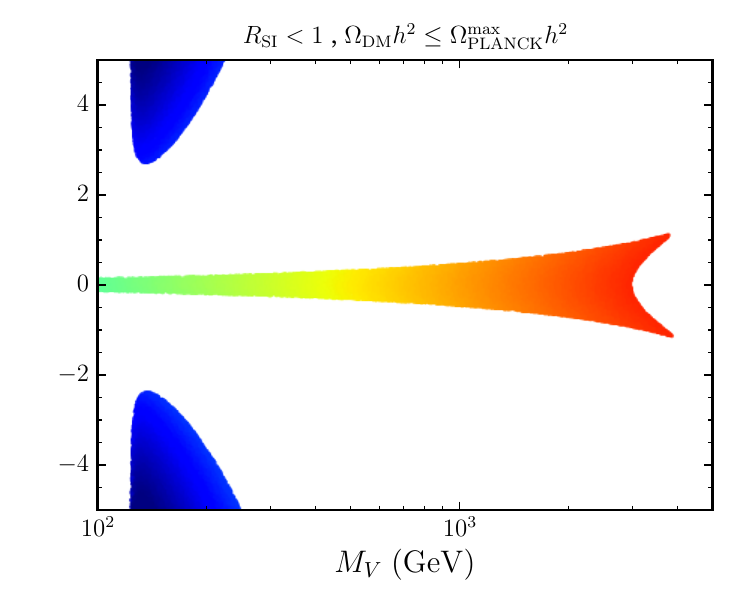}\hspace{-0.255cm}\includegraphics[height=0.25\textwidth]{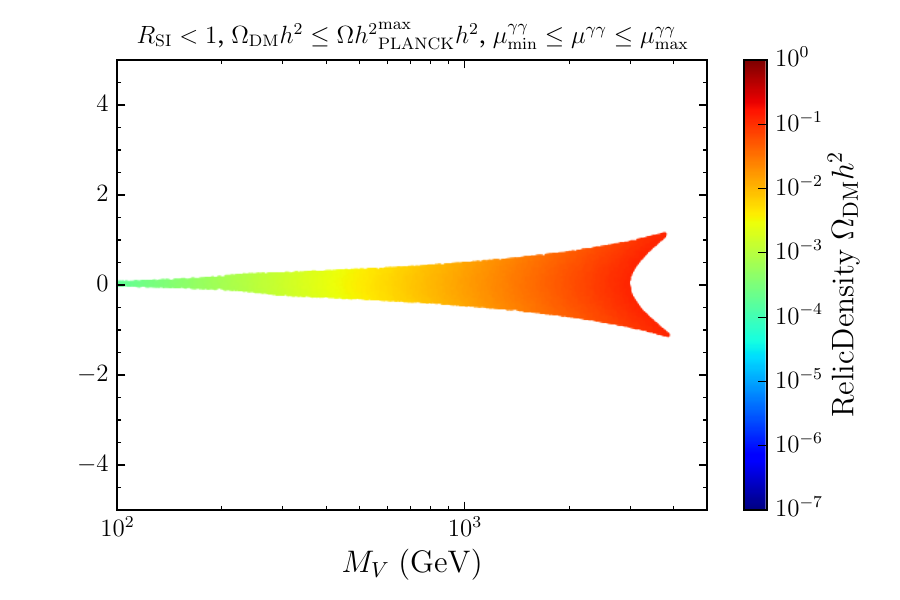}}
\caption{\label{fig:scan2} Impact of experimental bounds on the vector isotriplet parameter space.  We progressively impose the constraints from direct detection from XENON 1T 2018 (\textit{left}), the upper bound on the relic abundance (\textit{center}), and $R_{\gamma \gamma}$ (\textit{right}).  In Fig.~\ref{fig:scan3} we showed a focused plot on the remaining parameter space with the constraint of the relic abundance within the sigma of the Planck measurement.}
\end{figure}

\begin{figure}[htb]
\centering
\includegraphics[width=0.6\textwidth]{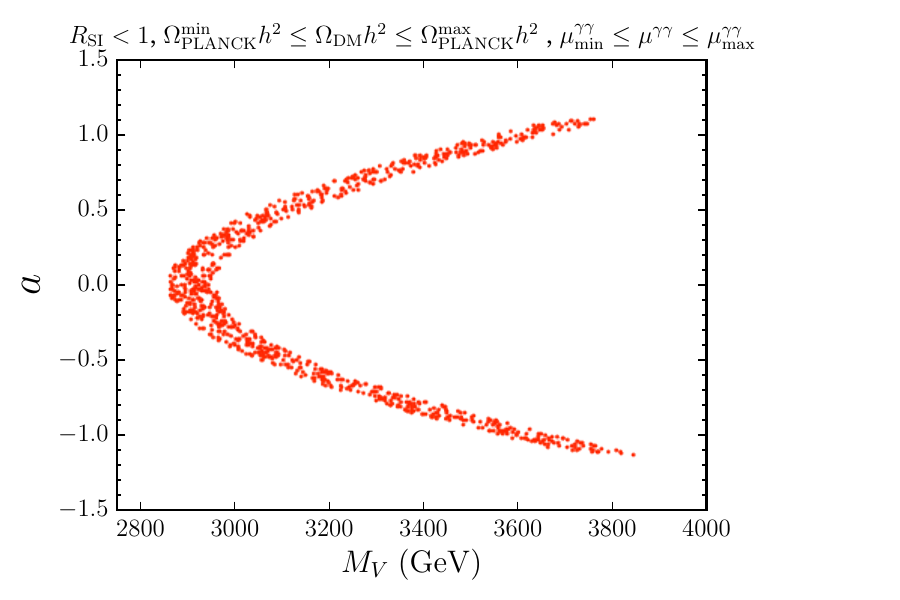}
\caption{\label{fig:scan3} The remaining parameter space in the vector isotriplet model with constraints imposed from direct detection, the Higgs coupling to photons, and the relic abundance.  For this case we demand that the relic abundance of vector DM is within one sigma of the Planck measured value.}
\end{figure}

Direct bounds on the lifetime of the charged component, $V^+$ ,can be obtained by LEP searches for long lived charginos~\cite{Heister:2002mn,Abbiendi:2002vz,Abdallah:2003xe} in scenarios with compressed spectra.  As the vector has an even larger production cross-section than a chargino, we can see that the bound reaches the maximal available mass (half the centre of mass energy of the collider). Thus, DM masses below $104\,$ GeV are excluded, which is in a range where the perturbative unitarity loss scale is also below our required limit.

In Fig.~\ref{fig:scan2} we show the effect of including constraints from direct detection, consideration of the relic density and measurement of the Higgs coupling to photons. The recent results from XENON1T rule out most of the parameter space, as shown in the left panel, leaving a strip around $a \approx 0$ and two low-relic-density lobes at masses below 1 TeV.  The upper bound on the thermal relic abundance eliminates the high mass values, as seen in the centre panel.  In the right panel we impose the 95\% CL bound from both ATLAS and CMS on the Higgs to diphoton rate, corresponding to the orange band in Fig.~\ref{fig:HtoAA}. The net effect is that this constraint removes two low mass lobes, while the band around $a\approx 0$ is reduced in width for $M_V \lesssim 200\,$GeV.  The combination of these constraints provides an upper bound on the Higgs-vector coupling of $|a| < 1.2$ at the maximal allowed mass $M_V \approx 3.85\,$TeV.  Lower masses down to $M_V > 104\,$GeV are allowed with smaller values of $a$ and subsequently a lower thermal relic density. 

 In Fig.~\ref{fig:scan3} we show the remaining parameter space with the additional constraint that the relic abundance of vector DM be within one sigma of the Planck measured value.  This requirement imposes a lower bound on the DM mass of $M_V \approx 2.85\,$TeV.  It is important to note that combined constraints establish upper and lower limits on both $a$ and $M_V$, creating the opportunity to  cover it at future experiments.   In Section \ref{sec:pheno} we will discuss how the remaining parameter space can be tested using the LHC and future high energy hadron colliders.

\section{Phenomenology at hadron colliders\label{sec:pheno}} 

\subsection{Decay width of $V^+$}

\begin{figure}[tb]
  \centering
  \includegraphics[scale=0.5]{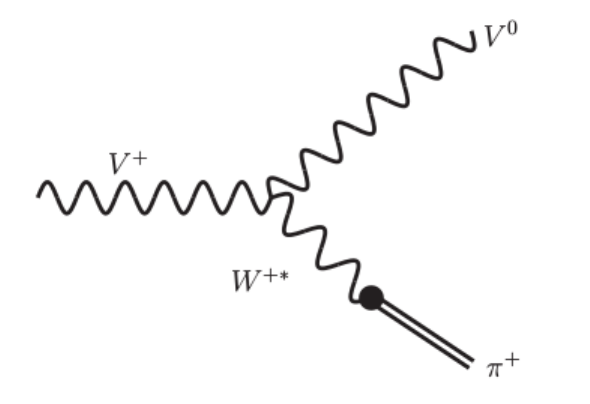}
  \caption{Feynman diagram of the effective $V^+ \to  V^0 \pi^+$ interaction, obtained by integrating out the off-shell $W$.}
  \label{fig:decay}
\end{figure}

Because of the mass splitting computed in Section \ref{sec:The-mass-splitting} the charged component of the vector isotriplet is short-lived before decaying into the neutral component plus hadrons or leptons. The main decay channel is the hadronic one, as the kinematically allowed leptonic channels ($V^+\rightarrow V^0 e^+ \nu_e$ and $V^+\rightarrow V^0 \mu^+ \nu_{\mu}$) produce partial widths which are at least one order of magnitude smaller.  As the small mass gap between  $V^+$ and $V^0$
is of the order of the pion mass, the naive perturbative
calculation of $V^+ \to  V^0 u\bar{d}$ fails to predict the correct 
width and lifetime of $V^+$.
For a proper evaluation of the lifetime, which is crucial for collider phenomenology,
we have used the effective Lagrangian for $V^+ \to  V^0 \pi^+$ interactions.
In momentum space this Lagrangian is
\begin{equation}
\mathcal{L}_{\pi^-V^+V^0}  =  \frac{g^2 f_{\pi}}{2\sqrt{2}M_W^2} 
[
(p_{V^+}-p_{V^0})_\alpha g_{\beta\gamma}  
+ {p_{V^0}}_\beta   g_{\alpha\gamma}
- {p_{V^+}}_\gamma  g_{\alpha\beta}
 ]p_{\pi^-}^\alpha \pi^-{V^+}^\beta {V^0}^\gamma,
\label{EFT-pion}
\end{equation}
where $p_i$ are the momenta of the respective particles and $f_{\pi}=130$ MeV is the usual pion decay constant. This effective Lagrangian is based on the diagram shown in Fig.~\ref{fig:decay}, where the virtual $W$ boson is integrated out and the effective $W-\pi$ mixing is described by the well-known coupling
\begin{equation}
  \mathcal{L}_{W\pi}  = \frac{g f_{\pi}}{2\sqrt{2}}W_{\mu}^{+}\partial^{\mu}\pi^{-} + 
\mathrm{h.c.}
\label{eq:wpi}
  \end{equation}

\begin{figure}[htb]
\centering
\includegraphics[width=0.5\textwidth]{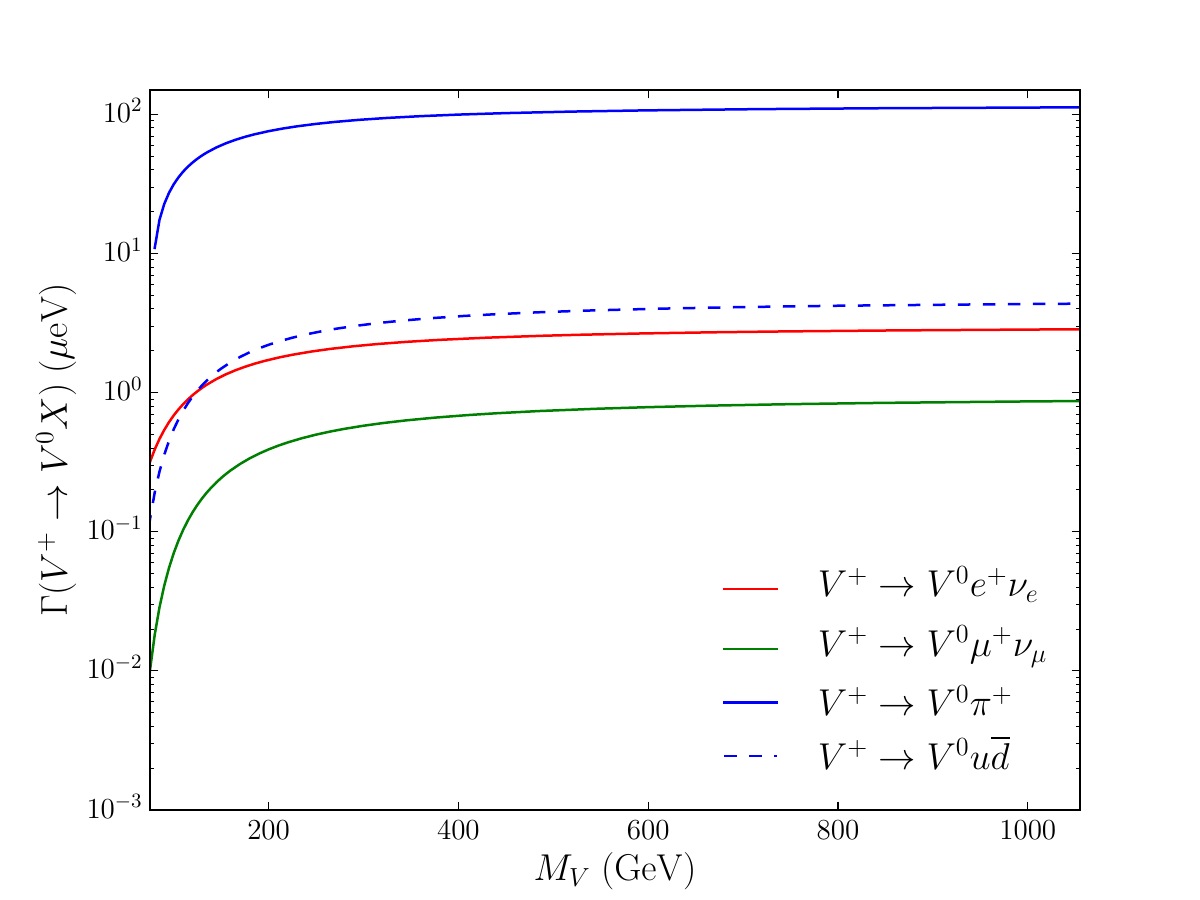}%
\includegraphics[width=0.5\textwidth]{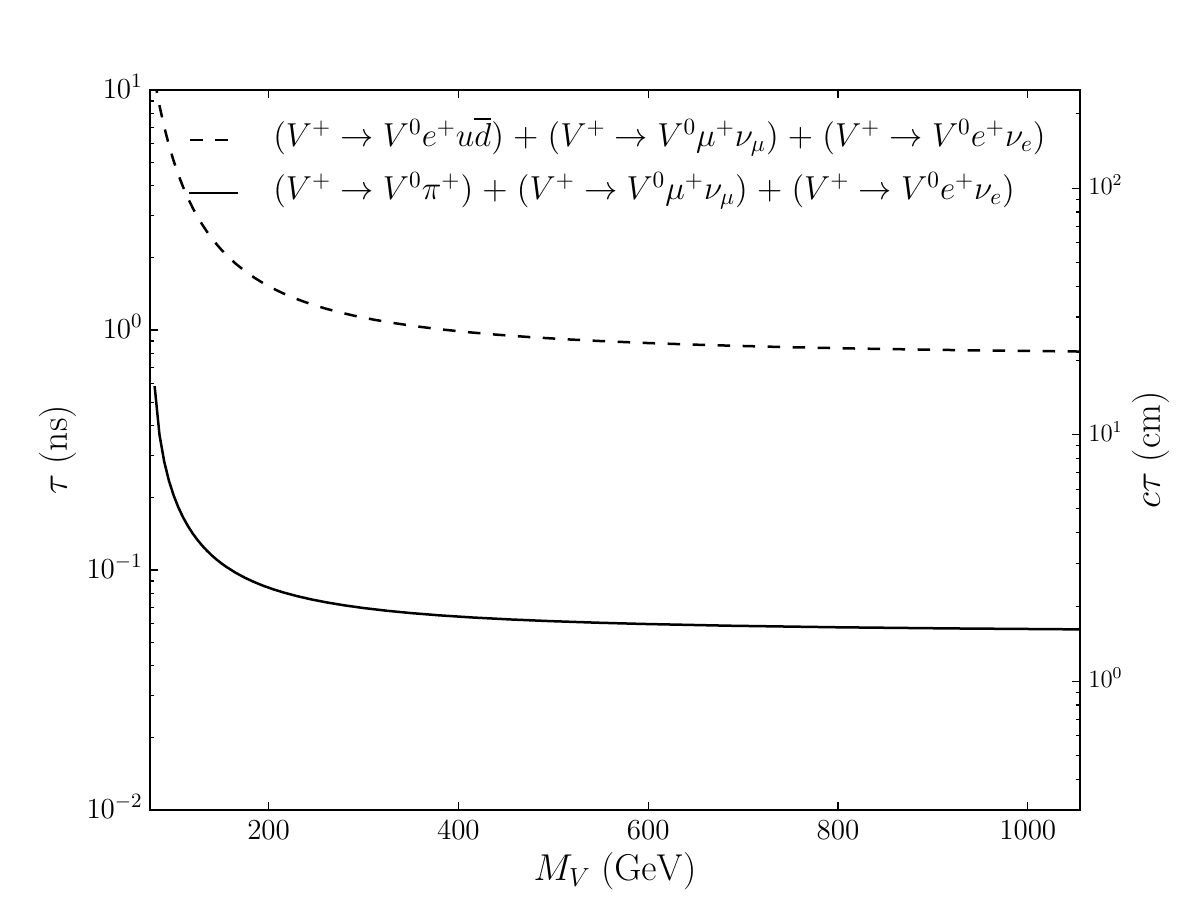}%
\caption{\label{fig:lifetime} \textit{Left:} The partial decay widths of the charged vector isotriplet component in $\mu$eV for the three allowed decay channels are given by sold lines. The dashed line indicates the naive perturbative result for the hadronic one. \textit{Right:}  The decay lifetime as a function of the mass, compared to the naive perturbative result.}
\end{figure}

In the left panel of Fig.~\ref{fig:lifetime} we compare the values of  the partial decay widths in the various channels, including the naive result for the hadronic one.  The decay widths, which are proportional to the mass splitting, tend a mass independent value in the same way as $\Delta M$ does in Fig.~\ref{fig:mass_splitting}.  We also note that the leptonic channels are sub-leading compared to the hadronic one, giving $\mbox{BR} (V^+ \to V^0 e^+ \nu_e) \approx 2.4\%$ and $\mbox{BR} (V^+ \to V^0 \mu^+ \nu_e) \approx 0.65\%$.  Finally we see that the naive calculation for the hadronic channel would predict a partial width roughly one order of magnitude smaller than the one obtained with the effective pion coupling. This results in a $V^+$ lifetime, shown in the right panel of Fig.~\ref{fig:lifetime}, of about $0.06$~ns, which corresponds to a mean decay length of about $2$~cm.

\subsection{Signal at hadronic colliders, LHC and FCC}

Due to the short decay lifetime of the charged isotriplet component most of the $V^\pm$ that would be produced at 
hadronic colliders would decay in a soft pion before hitting the tracker.  This would result in missing transverse energy. 
Therefore typical searches designed for DM should apply, such as monojet searches based on
a jet radiated by the initial state. However, the recoil against the jet will boost the produced $V^\pm$, thus 
allowing a sizeable number of these charged particles to decay inside the tracker. Therefore the most effective way to search for the
isotriplet is to look for charged tracks that disappear inside the tracker. This kind of search has
been performed by both ATLAS~\cite{Aaboud:2017mpt} and CMS~\cite{Sirunyan:2018ldc} during Run-I and Run-II, using the electroweak production of 
charginos in supersymmetry as a benchmark model. In the following, we will reinterpret the ATLAS search
for our model, as it is more sensitive to short lifetime charginos. A disappearing track is identified by the inner pixel tracker. An improvement in the ATLAS detector, i.e. the insertion of an inner 
layer of pixel trackers~\cite{Abbott:2018ikt} during the shutdown between Run-I and Run-II, allowed the identification of tracks disappearing between 12 and 30 cm, while the previous Run-I analyses were based on tracks disappearing
after 30 cm~\cite{Aad:2013yna}.  This improvement allows us search particles with shorter decay lifetimes, with the requirement that one jet has a transverse momentum above $140$~GeV.

\begin{figure}
\includegraphics[scale=0.7]{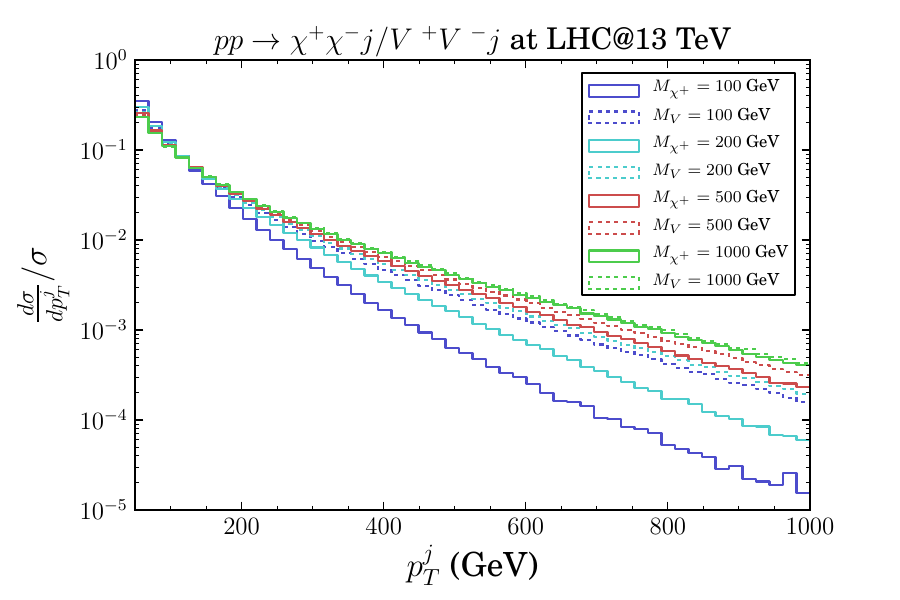}
\caption{\label{fig:jetpT} The shapes of $p_T$ distributions for pair production of the charged component of the vector isotriplet and of
a wino, for different values of the masses.}
\end{figure}

\begin{figure}
\includegraphics[scale=0.8]{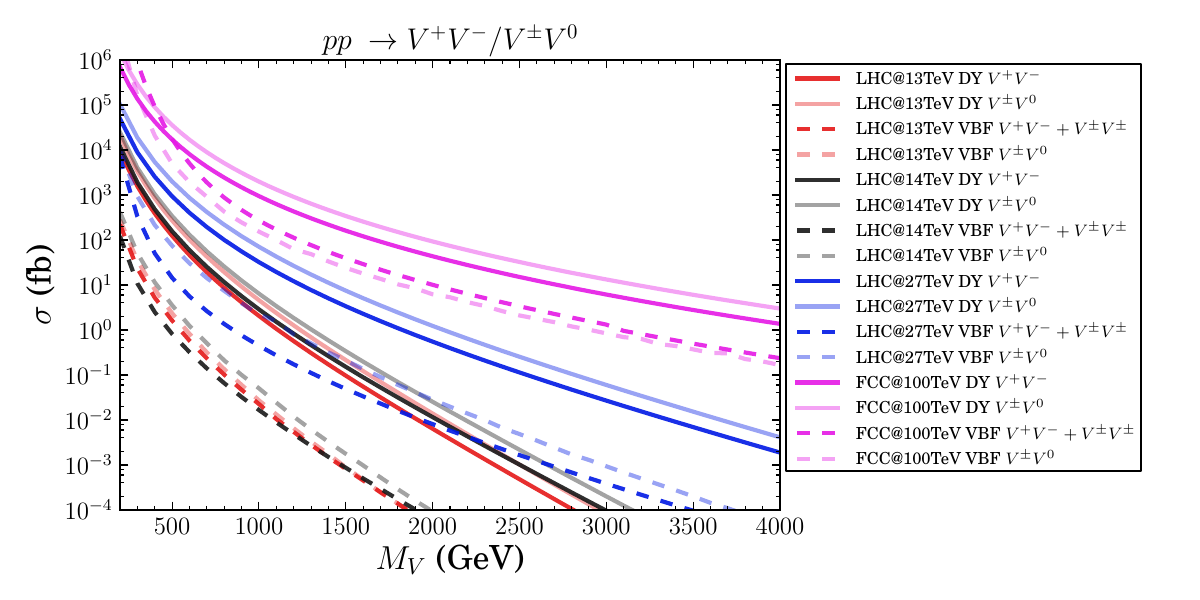}
\caption{\label{fig:xsecs} Production cross-sections at leading order for the vector isotriplet model
at various LHC energies and for a $100$~TeV future collider. The solid lines represent
Drell-Yann, while the dashed ones refer to the subleading vector boson fusion channel.
}
\end{figure}

\begin{figure}
\includegraphics[scale=0.8]{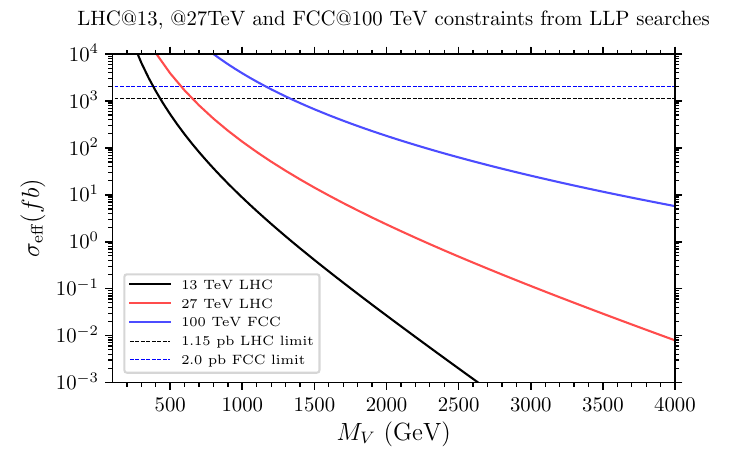}
\caption{\label{fig:limits} Effective cross-sections
$\sigma_{eff}=\sigma(pp\to V^\pm V^0)+2\sigma(pp\to V^+V^-)$
at leading order for the vector isotriplet model for 13 and 27 TeV LHC energies
 and for a $100$~TeV future collider. The dashed
 black line corresponds the current LHC sensitivity of about $0.85$~fb
 while dashed red line corresponds to the expected 2 fb sensitivity
 of 27 TeV LHC and 100 TeV FCC collider in high luminosity regime}
\end{figure}

We use an implementation of our model in CalcHEP\cite{Belyaev:2012qa}, using
LanHEP~\cite{Semenov:1996es,Semenov:1997qm,Semenov:2008jy}, to reinterpret these results.  The dominant production
channel for the vector isotriplet is the same as for 
winos, supersymmetric partners of W- and Z-bosons, namely electroweak Drell-Yann. As a first
test, we computed the jet $p_T$ distribution in the two models, as shown in Fig.~\ref{fig:jetpT}.
We see that the $p_T$ spectrum is always harder for the vector isotriplet model 
especially at masses below about $500\,$GeV for the LHC at $13\,$TeV.
This means that the fraction of $V^\pm$ that decays inside the tracker will be
greater for our model than for the ATLAS benchmark one. We thus decided to simply rescale
the ATLAS results to the cross-sections for the vector isotriplet, knowing that this  leads to a conservative
estimate of the bound on our model. As ATLAS only provides bounds on the cross-section after cuts,
we used our simulation to estimate the bound on the pair production cross-section for a mean
decay length of $2\,$cm for $\tau\simeq 0.06\,$ns.

In Fig.~\ref{fig:xsecs} we present production cross-sections at leading order for the vector isotriplet model at various LHC energies and for a $100\,$TeV future collider. The solid lines represent
Drell-Yann, while the dashed ones refer to the subleading vector boson fusion channel. 
In our evaluations we have used NNPDF23 LO \texttt{as\char`_0130\char`_QED}
\cite{Ball:2013hta}, and theQCD scale $Q$ is set to be the invariant mass of the final state particles, $\hat{s}$.
We can see that increasing the collider energy has a dramatic effect 
on the possibility to explore this model, as the cross-sections increase significantly for masses in the range of a few TeV.
For example, at a future $100\,$TeV hadron collider the cross-section 
of Drell-Yann production is always above the $1\,$fb level in the relevant mass range.

To estimate current and future collider sensitivities
to the model, we used the following procedure.
We have found that for $\tau\simeq 0.06\,$ns 
the current ATLAS limit for the wino mass is about $250\,$GeV~\cite{Aaboud:2017mpt}, which
corresponds to an effective cross-section 
$\sigma_{\text{eff}}=\sigma(pp\to V^\pm V^0)+2\sigma(pp\to V^+V^-)$
{about $1.15\,$pb.}  
By comparing this limit with the respective $\sigma_{\text{eff}}$
from our model, as shown by a black line in Fig.~\ref{fig:limits},
we estimate the current limit on the model to be 
{about 500 GeV}. 
We remark that this is a rather  impressive limit on the mass of a DM candidate when compared to other models.
The high-luminosity run of the
LHC may improve the bound, however it will not be able to push it in the interesting range of 
masses where the vector isotriplet can be the only DM candidate as the cross-sections above $2.8\,$TeV
remain too low.

To estimate the reach of a future high-energy collider, we follow the strategy delineated in Ref.~\cite{Mahbubani:2017gjh}, where
a projection is obtained by increasing the cut on the jet $p_T$ in order to match the number of
background events of the ATLAS search.\footnote{Ref.~\cite{Mahbubani:2017gjh} uses the results from the Run-I
analysis~\cite{Aad:2013yna}, but it also implements the Run-II improvement thus assuming reconstructed disappearing tracks
at a distance of 10 cm from the beamline.}  Using the final results of Ref.~\cite{Mahbubani:2017gjh}, we estimate 
that a missing track-based search at a 100 TeV collider will be able to exclude cross-sections in the
ballpark of {$2\,$pb} after an integrated Luminosity of $3000\,$fb$^{-1}$. From Fig.~\ref{fig:limits},
we see that this would imply a reach {about 1.2 TeV which is still not enough to probe the entire model}.
As a comparison, we also show the cross-sections for a high-energy
option of the LHC, using the reference value at $27\,$TeV~\cite{HL-LHC-WG}. Assuming that the 
{Assuming that the sensitivity to the disappearing charged tracks rate can be as good as for the $100\,$TeV case, it  would only enable us to probe masses up to 700 GeV.}

{We see  that even a $100\,$TeV future collider
	 would not be able to fully  probe  this model with  disappearing charged tracks -- the most promising signature so far, so one should explore some other means  to fully test this model.}

\section{Conclusions\label{sec:Conclusions}}

In this study we have constructed a minimal extension of the SM by introducing a new massive spin-one isotriplet field. To satisfy perturbative unitarity, this field needs to
be odd under a new $Z_{2}$ symmetry, making the neutral component
of the triplet a DM candidate. This minimal  isotriplet vector DM model introduces only two new parameters: the mass of the
DM particle and its coupling to the Higgs boson. We have shown that if the new particle has a mass in the range $2.8\,$TeV $\leq M_{V}\leq3.8\,$TeV the model can explain the measured DM relic density
and simultaneously satisfy a number of complementary current experimental constraints.
We have also shown that these constraints, ranging from direct
DM detection, to LEP bounds and the measurements of the
Higgs couplings to photons, also have an important interplay 
in setting an upper  limit on the absolute value of isotriplet coupling to the Higgs 
(which is currently about one)
and the isotriplet mass to be above $100\,$GeV. Masses above 
$3.8\,$TeV predict a DM density above the Planck measured value 
and are thus excluded by the over-closure of the Universe.

The most striking signal at hadron colliders is the presence of disappearing
charged tracks coming from the charged component of the vector isotriplet. We calculated
the lifetime of this charged component to be $\approx 0.06\,$ns.  A reinterpretation of current ATLAS and CMS searches at the LHC
allows us to exclude masses up to 
{about 500 GeV}. 
{It is worth to note that the exclusion limit we have established is conservative since the 
transverse momentum distribution of $V^\pm$ is harder than that from
charginos used in ATLAS analysis, so the respective efficiencies for disappearing
charged tracks from our vector DM model are higher, respectively leading to higher signal rate.}
We have also estimated that 
future hadron collider with a centre-of-mass energy of $100\,$TeV would be able to 
probe masses 
{up to about 1.2 TeV and therefore would not be able to fully  test  this model, so one should explore  further how to  cover the entire parameter space of the  model.}

\section*{Acknowledgments}

The authors thank P.~Schwaller and J.~Zurita for useful discussions and email exchanges on clarifications of the results in
Ref.~\cite{Mahbubani:2017gjh}. We would also like to thank A.~Pukhov, P.~Scott and I.~Shapiro for valuable discussions. This work
was partially financed by Fondecyt grants 1120346, 1160423, USM internal project USM PIIC program, Conicyt (Chile) grants ACT-1406
and PIA/Basal FB0821. GC acknowledges partial support from the Labex-LIO (Lyon Institute of Origins) under grant ANR-10-LABX-66 
(Agence Nationale de la Recherche) and FRAMA (FR3127, F\'ed\'eration de Recherche ``Andr\'e Marie Amp\`ere''). AZ is very thankful
to the developers of MAXIMA \cite{maxima} and the package Dirac2 \cite{Dirac2}. These software packages were used in parts of this
work. AB acknowledges partial  support from the STFC grant ST/L000296/1, Royal Society Leverhulme Trust Senior Research Fellowship
LT140094, and Soton-FAPESP grant. AB also thanks the NExT Institute and   Royal Society International Exchange grant IE150682,
partial support from the InvisiblesPlus RISE from the European Union Horizon 2020 research and innovation programme under the
Marie Sklodowska-Curie grant agreement No 690575.  JM was funded by the Imperial College London Presidents PhD Scholarship. 
TT and PM would like to thank FAPESP for support through grant 2013/01907-0. TT would  like also to thank FAPESP for support
through grant 2016/15897-4.

\appendix
\section{One-loop self-energies\label{sec:app}}

\subsubsection*{Definitions}
We define the one-loop integrals as
\begin{eqnarray}
\mathbf{A}(m) &=& 16\pi^2Q^{4-d}\int{d^dq\over i\,(2\pi)^d}{1\over
q^2+m^2}= -\frac{x}{\hat{\epsilon}} + A(x) + \mathcal{O}(\epsilon) \label{A0 def}\\
\mathbf{B}(m_1,m_2) &=& 16\pi^2Q^{4-d}\int{d^dq\over i\,(2\pi)^d}
{1\over\left[q^2+m^2_1\right]\left[
(q-p)^2+m_2^2\right]} = \frac{1}{\hat{\epsilon}}+B(x,y)+ \mathcal{O}(\epsilon) \ \ \label{B0 def}
\end{eqnarray}
where $d=4-2\epsilon$, $1/\hat{\epsilon} = 1/\epsilon-\gamma_\mathrm{E}+\log(4\pi)$ ($\gamma_\mathrm{E}$ is the Euler-Mascheroni constant) and $Q$ is the renormalisation scale.  For brevity we will write $\mathbf{B}(p, m_1,m_2) = \mathbf{B}(m_1,m_2)$ when the choice of the external momentum is clear.

We make use of the limiting case when $m_1=M\gg m_2$ and $p^2=M^2$ for the integral in Eq.~(\ref{B0 def})
\begin{equation}\label{eqn:B0_limit}
\mathbf{B}(M,M,m) = \frac{1}{\hat{\epsilon}}+ 2-\log\left(\frac{M^2}{\mu^2}\right)-\frac{\pi m}{M}+\mathcal{O}\left(\frac{m^2}{M^2}\right).
\end{equation}
The integral in Eq.~(\ref{A0 def}) is integrated to give
\begin{align}
\mathbf{A}(m)=m^2\left(\log\left(\frac{m^2}{Q^2}\right)-1+\frac{1}{\hat{\epsilon}}\right) \label{eqn:A0_def}
\end{align}
from which the limiting behaviour for large $m$ is clear.

\subsubsection*{Self-energies}
The one-loop self-energies of the charged and neutral components of the vector field $V$ are given by
\begin{eqnarray}
\Sigma_{V,Z}^+&=&\frac{19 g^2}{6 (16\pi^2)}\left[4-57\left(-c_W^2\mathbf{B}(M_Z,M_V)+\mathbf{B}(M_W,M_V)-\mathbf{B}(M_V,0)\right)\right] +\delta_Z\\
\Sigma_{V,M}^+ &=& \frac{g^2}{6(16\pi^2)}\left[ 16\left( \left(c_W^2M_V^2+M_W^2\right) \mathbf{B}(M_Z,M_V)+(M_V^2+M_W^2)\mathbf{B}(M_W,M_V) \right. \right.\nonumber\\
&& \left.\left.+M_V^2s_W^2\mathbf{B}(M_V,0)\right)+7c_W^2\mathbf{A}(M_Z)+14 \mathbf{A}(M_V) +7 \mathbf{A}(M_W) + 20 (M_V^2+M_W^2)\right]  \nonumber \\
&& \frac{a}{16\pi^2}\left(\mathbf{A}(M_H)+2\mathbf{A}(M_W)+\mathbf{A}(M_Z)\right) - \frac{a^2M_W^2}{\pi^2}\mathbf{B}(M_H,M_V)+ \delta_M \\
\Sigma_{V,Z}^0&=&\frac{g^2}{9 (16\pi^2)}\left[ 2 + 57 \mathbf{B}(M_W,M_V) \right]+\delta_Z\\
\Sigma_{V,M}^0 &=& \frac{g^2}{3(16\pi^2)}\left[ 7 \mathbf{A}(M_V) + 7 \mathbf{A}(M_W) + 2 \left( M_V^2+M_W^2\right)\left( 5 + 8\mathbf{B}(M_W,M_V)\right) \right] \nonumber \\
&& \frac{a}{16\pi^2}\left(\mathbf{A}(M_H)+2\mathbf{A}(M_W)+\mathbf{A}(M_Z)\right) - \frac{a^2M_W^2}{\pi^2}\mathbf{B}(M_H,M_V)+ \delta_M.
\end{eqnarray}

We check that the one-loop divergences are canceled by the corresponding counterterms.  The required counterterms are
\begin{eqnarray}
\delta_Z&=& \frac{19 g^2}{3(16 \pi ^2)\hat{\epsilon}} \\
\delta_M & = & \frac{a^2s^2_W}{\pi^2\hat{\epsilon}}M_W^2+\frac{a}{16\pi^2\hat{\epsilon}}\left(M_H^2+2M_W^2+M_W^2\right)- \frac{3g^2}{16\pi^2\hat{\epsilon}} \left[ M_V^2+M_W^2  \right].
   \end{eqnarray}

\subsubsection*{Series expansion of the mass splitting for large masses}

The one-loop mass splitting is given by the first term of series expansion in Eq.~(\ref{eqn:deltam}).   The series expansion can be represented as
\begin{equation}
\Delta M = 5 (M_W-c_W^2 M_Z)   \sum_{n=0}^{\infty}  \frac{(-1)^{n^2}c_n}{16^n} \left(  \frac{   g_W^{2(n+1)}  }{ \pi^{(2n+1)}   } \right) {\frac{1}{2} \choose n}  
 \left(42 \log\left(\frac{M^2}{Q^2}\right)-103\right)^n \label{eqn:deltam_series}
\end{equation}
where we have computed the coefficients $c_n$ up to $n=7$ in the limit of large $\Mv$ and verified that they are convergent.
\bibliography{bib}

\end{document}